\documentclass{article}

\usepackage{amsmath, amssymb, graphics, setspace}
\usepackage{float}
\usepackage{graphicx, color}

\begin{document}

\title{\LARGE {Current Algebra Formulation of} \\ 
       M-theory based on $E11$ Kac-Moody Algebra}
\author{{\sc Hirotaka Sugawara}\thanks{E-mail:{\ \  hirotaka.sugawara@oist.jp}.}}


\date{}
\maketitle

\vspace{-1.0cm} 
\begin{center}
{\it Okinawa Institute of Science and Technology, \\
1919-1 Tancha, Onna-son, Kunigami-gun, Okinawa  904-0495  JAPAN}
\end{center}
\vspace{1.9cm}

\begin{abstract}
Quantum M-theory is formulated using the current algebra technique.  
The current algebra is based on a Kac-Moody algebra rather than usual finite dimensional Lie algebra.  
Specifically, I study the $E_{11}$ Kac-Moody algebra that was shown recently\cite{West}
to contain all the ingredients of M-theory.  
Both the internal symmetry and the external Lorentz symmetry can be realized inside $E_{11}$, so that, by constructing the current algebra of $E_{11}$, I obtain both internal gauge theory and gravity theory.

The energy-momentum tensor is constructed as the bilinear form of the currents, yielding a system of quantum equations of motion of the currents/fields.  
Supersymmetry is incorporated in a natural way.  
The so-called ``field-current identity'' is built in and, for example, the gravitino field is itself a conserved super-current.  
One unanticipated outcome is that the quantum gravity equation is not identical to the one obtained from the Einstein-Hilbert action.
\end{abstract}

\section{Introduction}

Peter West and his coworkers\cite{West} have recently shown that a non-linear realization of the $E_{11}$ Kac-Moody algebra contains all the ingredients of M-theory.  
In particular, it contains 11-dimensional supergravity which is formulated in the normal Lagrangian scheme\cite{Cremmer}.   
\\
\\
Here I will present a quantum theory version of their theory.  
I perform the quantization using neither the customary path integral method nor Heisenberg's canonical commutation relation method.  
Instead, I employ the current algebra method that I developed some time ago\cite{HS}.  
The usual canonical commutation (or anti-commutation) relations pertain among extensive variables, but the current algebra method utilizes intensive variables -currents- to set up commutation relations.  
Therefore, the algebra is based on the symmetry of the physical system described by a Lie algebra or its extension - a Kac-Moody algebra.  
In fact, as I will show later in this paper, the idea of current-field identity\cite{Kroll} is built into our formalism, so that the commutation relations I adopt are both canonical and Lie algebraic (or rather Kac-Moody algebraic).  
\\
\\
Some examples of the field-current identity in our case are:  (1) the gravitino field integrated over space yields the supersymmetry generator; (2) the space integral of the spin connection provides the Lorentz generator in the flat tangent space; and, (3) the space integral of the elfbein field is the energy-momentum in the tangent space.  
\\
\\
Another important aspect of utilizing intensive variables is that I have no way to write down a local Lagrangian in terms of these variables.  
How, then, can I derive quantum equations of motion without the Lagrangian?  
This is where the Schwinger commutation relationcamong energy-momentum tensor comes in.  
Historically, Schwinger\cite{Schwinger} derived these relations in his 1963 papers assuming existence of a certain Lagrangian for the extensive variables.  
I assume his commutation relations are valid generally even when energy-momentum is written in terms of intensive variables such as currents, a well understood situation in two dimensional conformal theories such as string theory, where the Virasoro algebra is a two-dimensional version of the general Schwinger commutation relations.  
\\
\\
Schwinger noticed in his second 1963 paper that:  (1) the commutation relations are valid in curved space with some background metric; 
and (2) only the first derivative of the delta function appears if the fields are limited to spin 0, 1/2 and 1.  
He explicitly proves that the third derivative of the delta function appears if spin 3/2 is included, and comments that the commutation relation is not valid when the spin 2 particle -graviton- is present.  
With these facts in mind I assume that the 11-dimensional energy-momentum tensor $\Theta_{\mu \nu} (x)$ satisfies:
\begin{equation}
    [\Theta_{00} (x), \Theta_{00} (y)] = -i \{ \Theta_{0M} )x) + \Theta_{0M} (y) \} \partial_{M} \delta (\vec{x} - \vec{y}) \label{eq:1}
\end{equation}
where $M$ ranges from $1$ to $10$ and $\vec{x}$ represents the space components of the 11-dimensional vector $x$.
The commutation relation is valid at any point in the curved 11-dimensional space-time but I take the frame where the background metric is flat at that point $x$:  
\begin{equation*}
    g_{\mu \nu} (x) = \eta_{\mu \nu} = \mbox{diagonal } (-1, 1, \dots, 1).
\end{equation*}
The question then arises: how 
can the gravitational spin 2 field be incorporated assuming equation (\ref{eq:1}), which seems to be valid only for spin 0, 1/2 and 1?  
In fact, Schwinger asserts\cite{Schwinger} that I must add:
\begin{equation*}
    \partial_{k} \ \partial_{l} \ \partial'_{m} {\partial'}_{n} \delta (x-x') f^{kl, mn} (x),
\end{equation*}
to equation (\ref{eq:1}) when there is a spin 3/2 particle in the system.  
The exact two-dimensional 
case, the Virasoro algebra, is known to contain such a higher derivative term.
\\
\\
The answer is that my theory contains vector current only, and the gravitational field is a composite field of these vectors.  
Even in the ordinary formulation of gravity theory, one can write the metric tensor in term of the vierbein, a vector in curved space-time.  
In my case the relation is not merely technical but instead fundamental, in the sense that I do not assume the existence of a gravitational action written in terms of a metric tensor; the Spin 2 field is fundamentally composite.  
\\
\\
All the currents correspond to generators of $E_{11}$ or of its extension to be defined later.  
More precisely, the time component of each current, when integrated in 10-dimensional space, yields the $E_{11}$ generator.  
The energy-momentum tensor is expressed as the bilinear form of these currents and it is proven to satisfy the above Schwinger commutation relation.  
More precisely, 
the commutation relations among currents are constructed to satisfy the Schwinger commutation relation. 
Although for convenience I write the current commutation relations at a point where the metric is taken to be locally diagonal $\eta_{\mu \nu} = diagonal (-1, 1. ..., 1)$, 
they are valid in all of curved space-time.
\\
\\
The derived equations of motion are demonstrated to emerge in a generally covariant and locally Lorentz covariant way.
\\
\\
According to V. Kac\cite{Kac}, there are four ways to generalize the Lie group (algebra) to infinite set of generators:  group of diffeomorphism; current algebra; algebra of operators in Hilbert space; and the Kac-Moody algebra.  Mathematically speaking what I will be doing here is to combine two of the infinite extensions of the Lie algebra: current and Kac-Moody.  
In this sense I will be doubly extending the Lie algebra.  
The Kac-Moody part extends the symmetry of the physical system to an infinite dimensional algebra and the central extension (current algebra part) guarantees that the theory is a quantum field theory.
$E_{11}$ is a very extended algebra of the Lie algebra $E_{8}$, which is the largest gauge symmetry algebra encountered in some supergravity theories.
%
\begin{figure}[H]
   \centering
   \includegraphics[bb=0 0 642 105,width=12cm]{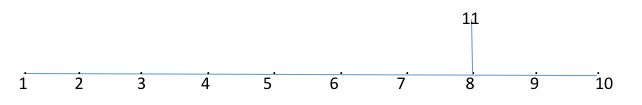}
   \caption{Dynkin diagram of $E_{11}$}
   \label{fig:1}
\end{figure}
\noindent
The Dynkin diagram of $E_{11}$ is shown in Figure 1 and it is obtained by lengthening the horizontal line of $E_8$ by 3 units.  
Lengthening by 1 unit yields the over extended algebra $E_{9}$, which is an affine algebra with vanishing determinant of the Cartan matrix.  
Lengthening by 2 units yields the very extended algebra $E_{10}$, which is Lorentzian with a negative value of the determinant of the Cartan matrix.  
[Over-extended algebra $E_{11}$ is given by lengthening by 3 units in the horizontal direction.]  
$E_{11}$ is the only algebra among  the above that which contains an obvious Lorentz algebra $O(10,1)$ as its subalgebra.  
Another way to look at the $E_{11}$ Dynkin diagram is to start from $A(11)$ diagram, which is a horizontal line with 10 points, and to add a vertical line at the 8th position.  
$A(11)$ contains $O(10,1)$ as its subalgebra.  
Serre's theorem\cite{Serre} shows that the multiple commutators of $E_{11}$ Chevalley generators given by the above Dynkin diagram do not finitely close thus leading to an (infinite) Kac-Moody algebra. 
\\
\\
P. West and his collaborators\cite{West} noticed that $E_{11}$ by itself does not contain all the ingredients needed for construction of M-theory.  
The most important missing element is the elfbein $e_{\mu}^{a}(x)$.  
They therefore extended the algebra by grading.  
The graded algebra is composed of $E_{11}$ and its vector representation $l (1)$.  
The commutation relation between the $l(1)$ components and the $E_{11}$ generators are dictated by the fact that $l (1)$ is a representation of $E_{11}$ together with the assumption that the commutation relations among the $l (1)$ components vanish.  
Thus a generator corresponding to $e_{\mu}^{a}(x)$ resides within $l (1)$.   
I need one more grading to super-symmetrize the theory.  
I assume there is an $E_{11}$ representation $sp (1)$ that behaves as a spinor representation under its subgroup $O(10,1)$.  
The commutation relation between the components of $sp (1)$ and the $E_{11}$ generators are determined by the representation matrix of $sp (1)$.  
The anti-commutation relation  among the $sp (1)$ components will be set up to be consistent with the supersymmetry.
\\
\\
The gravitino field $\psi_{\mu}^{\alpha} (x)$ corresponds to a  generator within $sp (1)$.  
\\
\\
P. West and his collaborators$^{(1)}$ depend heavily on a non-linear realization of the $E_{11}$ algebra; however, the non-linearity is unimportant to my development of a quantum version of their theory. 
\\
\\
For each generator $G^{a}$ of the algebra $E_{11} \times l (1) \times sp (1)$ there corresponds a current ${J^{a}}_{\mu} (x)$ that satisfies the commutation relation:
\begin{align}
    [J_{0}^{a} (x), J_{0}^{b} (y)] \mid_{x_{0}=y_{0}} &= if^{abc} J_{0}^{c} (y) \delta (\vec{x} - \vec{y}) \label{eq:2} \\
    [J_{0}^{a} (x), J_{N}^{b} (y)] \mid_{x_{0}=y_{0}} &= if^{abc} J_{N}^{c} (y) \delta (\vec{x} - \vec{y}) + iC \delta^{ab} \partial_{N} \delta (\vec{x} - \vec{y}) \label{eq:3} \\
    [J_{M}^{a} (x), J_{N}^{b} (y)] \mid_{x_{0}=y_{0}} &= 0 \mbox{ for } M, N \neq 0. \label{eq:4}
\end{align}
$f^{abc}$ is the structure constant of the algebra $E_{11} \times l (1) \times sp (1)$ and is given by
\begin{equation}
    [G^{a}, \ G^{b}] = if^{abc} G^{c} \tag{5} \label{eq:5}.
\end{equation}
When I consider the $sp (1)$ generators, the commutators among them must be changed to anti-commutators.  
The local commutation relations (\ref{eq:2}), (\ref{eq:3}) and (\ref{eq:4}) are assumed to be valid in the entire 11-dimensional curved space-time, but this particular form is valid at a point 
where the background metric is diagonal:  $\eta_{{\mu}{\nu}} = diag
(-1, 1, \cdots, 1)$.  
I observe that these commutation relations indicate not only the local symmetry of the system but also the quantum nature of the theory, especially equation (\ref{eq:3}) where the quantum nature is indicated by the derivative term.  
These commutation relations also imply a dual 	quality, extensive and intensive, for the current/field ${J^{a}}_{\mu} (x)$.  
The symmetry property is attributed to the intensive character and the quantum field property is described by the derivative term with constant $C$.  
\\
\\
I measure the space-time coordinate by some fundamental length 1 and so all the quantities appearing in equations (\ref{eq:2}), (\ref{eq:3}), (\ref{eq:4}) and (\ref{eq:5}) are dimensionless, including the constant $C$.  
One of the most important aspects of the above (anti)commutation relations is that they include a fermion (gravitino) in the system with the same value of the constant $C$.  
Note that usual Dirac particle yields a divergent term for this derivative term.  
\\
\\
I define
\begin{equation}
    \Omega_{\mu} (x) = J_{\mu}^{a} (x) G^{a} \tag{6} \label{eq:6}
\end{equation}
and write
\begin{align}
    \Theta_{\mu \nu} (x) &= \frac{1}{C} [tr (\Omega_{\mu} \Omega_{\nu}) - \frac{1}{2} \eta_{\mu \nu} tr (\Omega_{\rho} \Omega_{\rho}) ] \notag \\
    &= \frac{1}{C} [J_{\mu}^{a} (x) J_{\nu}^{a} (x) - \frac{1}{2} \eta_{\mu \nu} J_{\rho}^{a} (x) J_\rho^a (x)] \tag{7} \label{eq:7}.
\end{align}
\begin{equation*}
    tr (G^{a} G^{b}) = \delta^{ab} \mbox{ and } J_{\rho}^{a} (x) J_{\rho}^{a} (x) \mbox{ in fact means } \eta^{o \sigma} J_{\rho}^{a} (x) J_{\sigma}^{a} (x).
\end{equation*}
Then, as is shown in my old paper$^{(3)}$, the $\Theta_{{\mu}{\nu}}(x)$ satisfies the Schwinger commutation relation written in equation (\ref{eq:1}). 
\\
\\
For the purpose of a non-linear realization of the currents (inessential here), I can define, 
\begin{equation*}
g = \exp (\phi^{a}(x) G^{a})
\end{equation*}
and
\begin{equation}
    \Omega_{\mu} (x) = J_{\mu}^{a} (x) G^{a} = g^{-1} \partial_{M} \mbox{\Large $g$} \tag{8} \label{eq:8}.
\end{equation}
\\
$\phi^{a} (x)$ is the scalar field (tangent space spin 1/2 field in case of fermion) in the curved space-time and, with this non-linear realization, the entire theory can be described by the scalar and spinor field.  
\\
The quantum equations of motion for the current $J_{\mu}^{a} (x)$ are all derived from:
\begin{equation}
    -i \partial_{\mu} J_{\nu}^{a} (x) = [P_{\mu}, J_{\nu}^{a} (x)] \tag{9} \label{eq:9}
\end{equation}
with
\begin{equation}
    P_{\mu} = \int \Theta_{0 \mu} (x) \mathrm{d} \vec{x} 
\tag{10} \label{eq:10}.
\end{equation}
In passing, I should note that gravity is described by the vector field elfbein $e_{\mu}^{a} (x)$ and the connection fields $k_{\mu}^{ab} (x)$ rather than the metric field $g_{{\mu}{\nu}} (x)$ which is simply defined as a composite field:
\begin{equation}
    g_{\mu \nu} (x) = e_{\mu}^{a} (x) e_{\nu}^{a} (x) \tag{11} \label{eq:11}.
\end{equation}
\\
Our gravity theory is therefore fundamentally a vector theory in the curved space-time.

\section{$E_{11}$ algebra and its representations}

Here I summarize the minimum background on $E_{11}$ algebra required for later sections.  
Most of the equations in this section, except for the supersymmetry part, are found in the textbook by P. West\cite{West2}.  
First, the Dynkin diagram in Figure 1 gives the following Cartan matrix:
\begin{equation}
    A (E_{11}) = \begin{pmatrix}
                       A_{10} & 0 \\
                       & 0 \\
                       & 0 \\
                       & 0 \\
                       & 0 \\
                       & 0 \\
                       & 0 \\
                       & -1 \\
                       & 0 \\
                       & 0 \\
                       0, 0, 0, 0, 0, 0, 0, -1, 0, 0 & 2
                     \end{pmatrix} \tag{12} \label{eq:12}
\end{equation}
where $A_{10}$ is the Cartan matrix for the $A (10)$ (or $SU (11)$ or $SL (11)$) algebra with matrix elements 
\begin{equation*}
    a_{ii} = 2, \ \ a_{i, i+1} =a_{i+1, i} = -1 \mbox{ for } i = 1, \dots, 9.  
\end{equation*}
I have, therefore, 
\begin{equation}
    \det A (E_{11}) = -\det A (A_{7}) \det A (A_{2}) + 2 \det A (A_{10}) = -2 \tag{13} \label{eq:13}
\end{equation}
showing that $E_{11}$ is Lorentzian:  exactly one of the eigenvalues of Cartan matrix is negative.  
The above Cartan matrix gives the following system of simple roots $\alpha_{a}$ ($a = 1, \dots, r$; $r$ being the rank; $r = 11$ in this case), which is defined by the following condition on the inner products among simple root vectors:  
\begin{equation}
    (\alpha_{a}, \alpha_{b}) = A (E_{11})_{ab} \tag{14} \label{eq:14}.
\end{equation}
Each point (node) on the Dynkin diagram corresponds to one of the simple roots; all the points on the horizontal lines correspond to the simple roots of $A (10)$, which are given by:  
\begin{equation}
    \alpha_{a} = e_{a} - e_{a+1}, \ a =  1, \dots, 10 \tag{15} \label{eq:15}
\end{equation}
where $e_{a}$ constitutes an orthonormal system in the 11-dimensional Euclidean space.  
The last simple root corresponding to the [exceptional] node on top of the vertical line on the Dynkin diagram of Figure 1 is given by:
\begin{equation} 
    \alpha_{c} = - \lambda_{8} + x \tag{16} \label{eq:16}
\end{equation}
where $A (10)$ fundamental 
weight system \{$\lambda_{a}$\} is defined by:
\begin{equation} 
    (\alpha_{a}, \lambda_{b}) = \delta_{ab} \tag{17} \label{eq:17}
\end{equation}
with the solution:
\begin{equation}
    \lambda_{a} = \sum_{j=1}^a e_{j} - \frac{a}{11} \sum_{j=1}{11} e_{j} \tag{18} \label{eq:18}.
\end{equation}
\\
\\
I also get
\begin{equation}
    (A_{10}^{-1})_{ab} = (\lambda_{a}, \lambda_{b}) = \frac{a (11-b)}{11}; b \geq a \tag{19} \label{eq:19}.
\end{equation}
From (\ref{eq:16}) and (\ref{eq:18}) together with the definition of the Cartan matrix, I get
\begin{equation}
    2 = (\alpha_{c})^2 = (\lambda_{8} + x)^2 = 24 / 11 + x^2, \ x^2 = -2 / 11 \tag{20} \label{eq:20},
\end{equation}
indicating the Lorentzian nature of the $E_{11}$ root space.
\\
\\
For each simple root, I have three sets of Chevalley generators $H_{a}$, $E_{a}$, and $F_{a}$ satisfying:
\begin{align}
    [H_{a}, \ H_{b}] &= 0 \tag{21.1} \label{eq:21.1} \\
    [H_{a}, \ E_{b}] &= A (E_{11})_{ab} E_{b} \tag{21.2} \label{eq:21.2} \\
    [H_{a}, \ F_{b}] &= -A (E_{11})_{ab} F_{b} \tag{21.3} \label{eq:21.3} \\
    [E_{a}, \ F_{b}] &= \delta_{ab} H_{b} \tag{21.4}\label{eq:21.4}.
\end{align}
All the other generators of $E_{11}$ are obtained as multiple commutators of these Chevalley generators according to the Serre theorem, but the system does not close and the number of generators is infinite.
         
\subsection*{Some other  generators}

Some other generators than Chevalley's can in principle be obtained by multiply commuting the Chevalley ones but it is not practical.  
I must take a different path.
First, all the other generators correspond to either positive or negative roots are defined by
\begin{equation}
    \pm \alpha (l, m) = l (\lambda_{8} + x) + \sum_{j=1}^{r-1} m_{j} \alpha_{j} \tag{22} \label{eq:22}.
\end{equation}
Here, $r$ is the rank of $A_{10}$, namely $r=11$.  
The whole set of generators gives the adjoint representation of $E_{11}$, and the weight system of the adjoint representation is also given by equation (\ref{eq:22}) which contains not just the adjoint representation of $A(10)$ but other representations.  
Suppose that the highest weight representation with the highest (lowest) weight given by 
\begin{equation}
    \Lambda = \sum_{a=1}^{10} p_{a} \lambda_{a} \tag{23} \label{eq:23}
\end{equation}
is included in the adjoint representation.  Then I obtain
\begin{equation}
    \sum_{a=1}^{10} p_{a} \lambda_{a} = l (\lambda_{8} + x) + \sum_{a=1}^{r-1} m_{a} \alpha_{a} \tag{24} \label{eq:24}.
\end{equation}
This leads to
\begin{equation}
    \sum p_{a} (\lambda_{a} \lambda_{b}) = (\lambda_{8}, \lambda_{b}) + m_{b} \tag{25} \label{eq:25}.
\end{equation}
With equation (\ref{eq:19}), I get, switching the notation from $(a, b)$ to $(i, j)$,
\begin{gather}
    \sum_{i < j} ip_{i} (11-j) + \sum_{i \geq j} jp_{i} (11-i) = -8 l (11 - j) + 11 m_{j} \mbox{ for } 8 \leq j \tag{26} \label{eq:26} \\
    \sum_{i < j} ip_{i} (11 - j) + \sum_{i \geq j} jp_{i} (11-i) = -3 jl + 11 m_{j} \mbox{ for } 8 \geq j \tag{27} \label{eq:27}.
\end{gather}
The right hand sidehas opposite sign for the negative highest weight.
I investigate these constraints for the lowest weight.
For each level $l = 1, 2, 3, \dots$, I can write equations:
\\
(0) level 0 i.e. $l = 0$ case:

All the generators of $A(10)$ algebra. 
\\
(1) level 1 i.e. $l = 1$ case:

For $j=1$, I have
\begin{equation}
    \sum_{i \geq 1} p_{i} (11 - i) = 3 - 11 m_{1} \tag{28.1} \label{eq:28.1}.
\end{equation}

For $j=10$, I get
\begin{equation}
    \sum_{i \leq 10} ip_{i} = 8 - 11 m_{10} \tag{28.2} \label{eq:28.2}.
\end{equation}

By adding these equations I get
\begin{equation}
    \sum_{i \leq 10} p_{i} = 1 - (m_{10} + m_{1}) \tag{28.3} \label{eq:28.3}.
\end{equation}

The only solution is
\begin{equation}
    p_{8}=1 \mbox{ and all the other } p_{i}=0 \mbox{ and } m_{10}=m_{1}=0 \tag{28.4} \label{eq.28.4}.
\end{equation}
(2) level 2 i.e. $l = 2$ case:

For $j=1$,
\begin{equation}
    \sum_{i \geq 1} p_{i} (11 - i) = 6 - 11 m_{1} \tag{29.1} \label{eq:29.1}.
\end{equation}

For $j=10$,
\begin{equation}
    \sum_{i \leq 10} ip_{i} = 16-11 m_{10} \tag{29.2} \label{eq:29.2}.
\end{equation}

By adding these equations, I get
\begin{equation}
    \sum_{i \geq 1} p_{i} = 2 - (m_{1} + m_{10}) \tag{29.3} \label{eq:29.3}.
\end{equation}

The only solution is
\begin{equation}
    m_{1}=0, \ m_{10}=1, \ p_{5}=1 \tag{29.4} \label{eq:29.4}.
\end{equation}
(3) level 3:

Similarly I get as the only solution
\begin{equation}
    p_{2}=p_{3}=p_{10}=1 \tag{30} \label{eq:30}.
\end{equation}
In tensor notation, the $A(10)$ representation with Dynkin index $p_{i}=1$ corresponds to the antisymmetric tensor $T^{11-i}$, thus the level 1 generator is the 3rd rank antisymmetric tensor $T^{abc}$ and the level 2 generator is $T^{abcdef}$.  
The level 3 generator is $T^{abcdefgh, i}$ with the constraint 
\begin{equation*}
    T^{abcdefgh, \ i}=0.
\end{equation*}
I list all the commutators up to level 3.

Level 0 generators:
\begin{equation}
    [K_{b}^{a}, \ K_{d}^{c}] = \delta_{b}^{c} K_{d}^{a} - \delta_{d}^{a} K_{b}^{c} \tag{31.1} \label{eq:31.1}.
\end{equation}

Level 1 generators:
\begin{equation}
    [K_{b}^{a}, \ R^{cde}] = \delta_{b}^{c} R^{ade} + \delta_{b}^{d} R^{aec} + \delta_{b}^{e} R^{acd} = 3 \ \delta_{b}^{[c} R^{ade]} \tag{31.2} \label{31.2}.
\end{equation}

Level 2 generators:
\begin{equation}
    [K_{b}^{a}, \ R^{cdefgh}] = 6 \ \delta_{b}^{[c} R^{adefgh]} \tag{31.3} \label{eq:31.3}.
\end{equation}

Level 3 generators:
\begin{equation}
    [K_{b}^{a}, \ R^{cdefghij, k}] = 8 \ \delta_{b}^{[c} R^{adefghij], k} + \delta_{b}^{k} R^{cdefghij, a} \tag{31.4} \label{eq:31.4}.
\end{equation}
Commutators of level 1 generators yield level 2 generators:
\begin{equation}
    [R^{abc}, \ R^{def}]=2 R^{abcdef} \tag{31.5} \label{eq:31.5},
\end{equation}
and level 1 and 2 commutators will give level 3 generators:
\begin{equation}
    [R^{abc}, \ R^{defghi}]=-3 R^{defghi[abc]} \tag{31.6} \label{eq:31.6}.
\end{equation}
Positive and negative generators have commutators:
\begin{equation}
    [R^{abc}, \ R_{def}] = 18 \ \delta_{[de}^{[ab} K_{f]}^{c]} - 2 \ \delta_{def}^{abc} D \tag{31.7} \label{eq:31.7}
\end{equation}
where
\begin{equation}
    D = \sum_{b=1}^{11} K_{b}^{b}, \delta_{de}^{ab} = \frac{1}{2} [\delta_{d}^{a} \delta_{e}^{b} - \delta_{d}^{b} \delta_{e}^{a}] \tag{31.8} \label{eq:31.8}.
\end{equation}

\subsection*{Cartan involution invariant subgroup of $E_{11}$}

Corresponding to the Cartan involution:
\begin{equation}
    E_{a} \longrightarrow -\eta_{aa} F_{a}, \ F_{a} \longrightarrow -\eta_{aa} E_{a} \ H_{a} \longrightarrow -H_{a} \tag{32} \label{eq:32},
\end{equation}
I have the invariant subalgebra composed of the following generators:
\begin{equation}
    K_{ab} = K_{b}^{c} \eta_{ac} - K_{a}^{c} \eta_{bc} \tag{33} \label{eq:33}.
\end{equation}
This is either $SO(11)$ or $SO(10, 1)$ depending on whether $\eta$ is Euclidean or Lorentzian.  
I assume Lorentzian for the obvious reason.
\\
\\
Other generators include
\begin{equation}
    S_{abc} = R^{def} \eta_{da} \eta_{eb} \eta_{fc} - R_{abc} \tag{34} \label{eq:34}.
\end{equation}
These satisfy the following commutation relations:
\begin{align}
    [K^{ab}, \ K^{cd}] &= \eta^{bc} K^{ad} - \eta^{da} K^{cb} - \eta^{bd} K^{ac} + \eta^{ca} K^{db} \tag{35.1} \label{eq:35.1} \\
    [K^{ab}, \ S^{cde}] &= (\eta^{bc} S^{ade} + \eta^{bd} S^{aec} + \eta^{be} S^{acd}) - (a \longleftrightarrow b) \tag{35.2} \label{eq:35.2} \\
    [S^{abc}, \ S_{def}] &= -18 \ \delta_{[de}^{[ab} \ K_{f]}^{c]} S^{abc} \tag{35.3} \label{eq:35.3}.
\end{align}
In equation (\ref{eq:35.3}), $K_{f}^{c}$ must be understood as $\eta_{fg} K^{cg}$ rather than the original $K_{f}^{c}$ which is not Cartan reflection invariant.  
In this paper I concentrate on a few low level generators since the purpose of the work is to demonstrate the general methods of the current algebra formulation with some essential components to describe the gravity and the gauge theory that originate in M-theory.  
Thus I consider only the level 0 operators $K_{ab}$, which constitute the adjoint representation of the $A(10)$ algebra; and the level 1 generator $T^{abc}$, which seems to appear in the Lagrangian formulation of 11-dimensional supergravity\cite{Cremmer}.   
From the physics  point of view I will use the Cartan reflection invariant part $K_{ab}$ to define the gravitational connection field, thus making the connection a genuine gauge field.  
I need one more gauge field, the elfbein, to describe 11-dimensional gravity; the corresponding generator is not found among the above operators.  
The authors of reference (1) found this operator in the $l (1)$ representation of $E_{11}$, as will be described bellow.  
Physically speaking, this operator can be characterized as the energy-momentum operator in the 11-dimensional tangent space.  
Thus the elfbein is the gauge field corresponding to the tangent-space energy-momentum.  
To describe 11-dimensional supergravity, I need one more grading of $E_{11}$. 
The representation must include the 11-dimensional spinor representation because the supersymmetry operator belongs to the spinor representation.  
Corresponding to this spinor operator, I have a gauge vector field in the curved space-time; this is none other than the gravitino field.

\subsection*{l(1) representation}

The fundamental weight of $E_{11}$ can be written as:
\begin{equation}
    \Lambda_{j} = \lambda_{j} + \frac{x}{x^2} (\lambda_{8}, \lambda_{j}), \ \Lambda_{11} = x / x^{2} \tag{36} \label{eq:36}
\end{equation}
using 
\begin{align}
     (\lambda_{8}, \lambda_{1}) &= (\sum_{j=1}^{8} e_{j} - \frac{8}{11} \sum_{j=1}^{11} e_{j}, e_{1} - \frac{1}{11} \sum_{j=1}^{11} e_{j})  \notag \\
    &= 1 - 8 / 11 - 8 / 11 + 8 / 11  \notag \\
    &= \frac{3}{11}  \notag
\end{align}
and  
\begin{equation*}
    x^{2}=-2/11, 
\end{equation*}
I have 
\begin{equation}
    \Lambda_{1} = \lambda_{1} - \frac{3}{2} x \tag{37} \label{eq:37}.
\end{equation}
$l (1)$ representation is defined to be the highest weight representation with highest weight $\Lambda_{1}$.  
Corresponding to the weights $\Lambda_{1}$ I have the tensor $P^{a_{1}, \dots, a_{10}}$.  
Define
\begin{equation}
    P_{a} = \frac{1}{10} \epsilon_{a_{1} \dots a_{10}} p^{a_{1} \dots a_{10}} \tag{38} \label{eq:38}
\end{equation}
This operator can be called the energy momentum in the tangent space.  
The corresponding gauge field must be the vector in both the curved space-time and in the tangent space.  
It must be the elfbein field for the 11-dimensional gravity.
\\
\\
I can determine some other operators in $l (1)$ as follows, obtaining weights successively as
 \begin{gather}
    \mid \Lambda_{1} \rangle, \ \mid \Lambda_{1} - \alpha_{1} \rangle, \ \mid \Lambda_{1} - \alpha_{1} - \alpha_{2} \rangle, \ \mid \Lambda_{1} - \alpha_{1} - \dots - \alpha_{8} - \alpha_{9} \rangle, \notag \\
    \dots, \ \mid \Lambda_{1} - \alpha_{1} - \dots -\alpha_{8} - \alpha_{11} \rangle \tag{39} \label{eq:39}.
\end{gather}
The first 9 weights in the above series belong to the same representation of $A(10)$ but not the last one:
\begin{align}
    \Lambda_{1} -  \alpha_{1} &- \dots - \alpha_{8} - \alpha_{11} \notag \\
    &= \lambda_{1} - \frac{3}{2} x - \alpha_{1} - \dots - \alpha_{8} - (- \lambda_{8} + x) \notag \\
    &= \lambda_{1} + \lambda_{8} - \alpha_{1} - \dots - \alpha_{8} - \frac{5}{2} x \notag \\
    &= e_{1} - \frac{1}{11} \sum_{j=1}^{11} e_{j} + \sum_{j=1}^{8} e_{j} - \frac{8}{11} \sum_{j=1}^{11} e_{j} - e_{1} + e_{9} \notag \\
    &= \lambda_{9} - \frac{5}{2} x \tag{40} \label{eq:40}.
\end{align}
$\Lambda_{9}$ corresponds to the second rank tensor $Z^{ab}$.  
\\
Next, 
\begin{align}
    \lambda_{9} - \alpha_{11} &= \lambda_{9} + \lambda_{8} - \frac{7}{2} x  \notag \\
    &= \sum_{j=1}^{9} e_{j} - \frac{9}{11} \sum_{j=1}^{11} e_{j} + \sum_{j=1}^{8} e_{j} - \frac{8}{11} \sum_{j=1}^{11} e_{j} - \frac{7}{2} x  \notag \\
    &= \sum_{j=1}^{6} e_{j} - \frac{6}{11} \sum_{j=1}^{11} e_{j} + e_{7+8} - e_{10+11} - \frac{7}{2} x \tag{41} \label{eq:41}.
\end{align}
I also have,
\begin{equation}
    e_{7+8} - e_{10+11} = \alpha_{7} + 2 \alpha_{8} + 2 \alpha_{9} + \alpha_{10} \tag{42} \label{eq:42}.
\end{equation}
Subtracting this from (\ref{eq:41}), I get,
\begin{equation}
    \lambda_{9} - (\alpha_{7} + 2 \alpha_{8} + 2 \alpha_{9} + \alpha_{10}) - \alpha_{11} = \lambda_{6} - \frac{7}{2} x \tag{43} \label{eq:43}.
\end{equation}
This $\lambda_{6}$ corresponds to $Z^{abcde}$.  
I can go one more step further in the process:
\begin{equation}
    \lambda_{6} + \lambda_{8} - \frac{9}{2} x = \sum_{j=1}^{6} e_{j} - \frac{6}{11} \sum_{j=1}^{11} e_{j} + \sum_{j=1}^{8} e_{j}- \frac{8}{11} \sum_{j=1}^{11} e_{j} - \frac{9}{2} x \tag{44} \label{eq:44}.
\end{equation}
This can be either
(1)
\begin{equation}
    \sum_{j=1}^{4} e_{j} - \frac{4}{11} \sum_{j=1}^{11} e_{j} + \sum_{j=1}^{10} e_{j} - \frac{10}{11} \sum_{j=1}^{11} e_{j} + (e_{5+6} - e_{9+10}) - \frac{9}{2} x \tag{45} \label{eq:45}
\end{equation}
or
\begin{equation}
    \lambda_{3} + \alpha_{4} + 2 \alpha_{5} + 3 \alpha_{6} + 3 \alpha_{7} + 3 \alpha_{8} + 2 \alpha_{9} + \alpha_{10} - \frac{9}{2} x \tag{46} \label{eq:46}.
\end{equation}
The former corresponds to $Z^{abcdefg, h}$ and the latter to $Z^{abcdefgh}$.
One more step:
\begin{equation}
    \lambda_{3} + \lambda_{8} - \frac{11}{2} x = \sum_{j=1}^{3} e_{j} - \frac{3}{11} \sum_{j=1}^{11} e_{j} + \sum_{j=1}^{8} e_{j} - \frac{8}{11} \sum_{j=1}^{11} e_{j} - \frac{11}{2} x \tag{47} \label{eq:47}
\end{equation}
yields the regular representation.  
The Lie algebra $E_{11}$ can be graded as follows:  
Corresponding to the representation  
\begin{equation}
    u (A) \mid X_{s} \rangle = -D (A)_{st} \mid X_{1} \rangle \tag{48} \label{eq:48},
\end{equation}
I define the operators
\begin{equation}
    [X_{s}, \ A] = -D(A)_{st} X_{t} \tag{49} \label{eq:49},
\end{equation}
and, in this particular case of $l (1)$, I assume:
\begin{equation}
    [X_{s}, \ X_{t}] = 0 \tag{50} \label{eq:50}.
\end{equation}
I thus consider $E_{11} \oplus l_{1}$.
\\
The commutators are:
\begin{align}
    [K_{a}^{b}, \ P_{c}] &= - \delta_{c}^{b} P_{a} + \delta_{a}^{b} P_{c} \tag{51} \label{eq:51} \\
    [K_{a}^{b}, \ Z^{cd}] &= 2 \ \delta_{a}^{[c} \ Z^{bd]} + 1 / 2 \ \delta_{a}^{b} Z^{cd} \tag{52} \label{eq:52} \\
    [K_{a}^{b}, \ Z^{cdefg}] &= 5 \ \delta_{a}^{[c} \ Z^{bdefg]} + 1 / 2 \ \delta_{a}^{b} Z^{cdefg} \tag{53} \label{eq:53}.
\end{align}
Equation (\ref{eq:51}) gives  
\begin{equation}
    [K_{ab}, \ P_{c}] = - \eta_{bc} P_{a} + \eta_{ac} P_{b} \tag{54} \label{eq:54}.
\end{equation}
Of the $l (1)$ generators, in this paper I will be concerned only with $P_{a}$.

\subsection*{Spinor representation and the supersymmetry}

I do not have much understanding of the spinor representation of $E_{11}$.  
Here, I simply assume its existence and compute the necessary commutation relation in applying it to physics.  
One of the $A(10)$ representations of $E_{11}$ spinor representation is denoted as $\Psi$ and satisfies the following commutation relation with Cartan reflection invariant $O(10,1)$ generators, $K_{ab}$:
\begin{equation}
    [K_{ab}, \ \Psi^{\alpha}] = {\chi^\alpha}_{\beta, ab} \Psi^{\beta}, \ \ [S_{abc}, \ \Psi^{\alpha}] = {\zeta^{\alpha}}_{\beta, abc} \Psi^{\beta} \tag{55} \label{eq:55}.
\end{equation}
The structure constants $\chi_{\beta, ab}^{\alpha}$ will be calculated in the Appendix.  
The calculation of the constants $\zeta_{\beta, abc}^\alpha$ is not done in this paper.  
\\
$\Psi^{\alpha}$'s are Fermionic operators and obey anti-commutation relations: 
\begin{equation}
    \{ \Psi^{\alpha}, \ \overline{\Psi}^{\beta} \} = -2 P_{a} \Gamma^{\alpha \beta, a} + \xi^{\alpha \beta, abc} S_{abc} + \dots \tag{56} \label{eq:56}.
\end{equation}
The first term looks like usual 11-dimensional supersymmetry algebra and the second term indicates a possible ``central charge."  
The most important difference from the usual supersymmetry algebra is the energy-momentum operator $P_{a}$ rather than the energy-momentum in curved space-time given in equation (\ref{eq:7}) as the bilinear of the currents.  
In our formulation, the supersymmentry algebra is a subalgebra of the algebra:
\begin {equation}
    E_{11} \otimes l(1) \otimes sp(1) \tag{57} \label{eq:57}
\end{equation}
where $sp(1)$ is a spinor representation of $E_{11}$.

\section{Derivation of the quantum equation of motion}

I am now ready to write down the equation of motion for the currents of $E_{11} \otimes l (1) \otimes sp (1)$ algebra.  
I restrict myself in this paper to four types of currents corresponding to $K^{ab}$, $P^{a}$, $S^{abc}$ and $\Psi^{a}$ .  
The first two describe the 11-dimensional gravity: $S^{abc}$ is the anti-symmetric tensor needed to describe the 11-dimensional supergravity, and $\Psi^{\alpha}$ for the supersymmetry.
\\
In this approximation, I have
\begin{align}
    \Omega_{\mu} (x) &= J_{\mu}^{a} (x) G^{a} = g^{-1} \partial_{M} g \notag \\
    &= k_{\mu}^{ab} (x) K_{ab} + e_{\mu}^{a} (x) P_{a} + B_{\mu}^{abc} (x) S_{abc} + \psi_{\mu}^{\alpha} (x) \Psi_{\alpha} + \dots \tag{58} \label{eq:58}.
\end{align}
The necessary local commutation relations can be obtained by using the general formula equations (\ref{eq:2}), (\ref{eq:3}), (\ref{eq:4}), and (\ref{eq:5}), together with all the symmetry algebra derived in the previous section.
\\
I get, for the commutators that include at least one $e_{\mu}^{a} (x)$:
\begin{align}
    [k_{0}^{ab} (x), \ e_{\mu}^{c} (y)] |_{x_{0} = y_{0}} &= (- \eta^{ac} e_{\mu}^{b} (y) + \eta^{bc} e_{\mu}^{a} (y)) \delta (\vec{x} - \vec{y}), \notag \\
    &\ \ \ \mbox{ for } \mu = 0, 1, \dots, 9, 10 \tag{59.1} \label{59.1}, \\
    [k_{M}^{ab} (x), \ e_{N}^{b} (y)] |_{x_{0} = y_{0}} &= 0 \mbox{ for } M \neq 0 \tag{59.2} \label{59.2}, \\
    [e_{0}^{a} (x), \ e_{N}^{b} (y)] |_{x_{0} = y_{0}} &= i C \eta^{ab} \partial_{N} \delta (\vec{x} - \vec{y}) \mbox{ for } N \neq 0 \tag{59.3} \label{59.3}, \\
    [e_{0}^{a} (x), \ e_{0}^{b} (y)] |_{x_{0} = y_{0}} &= [e_{M}^{a} (x), \ e_{N}^{b} (y)] \mid_{x_{0} = y_{0}} = 0 \mbox{ for } M, N \neq 0 \tag{59.4} \label{59.4}.
\end{align}
%
Hereafter the capital Roman letters $M, N, \dots$ range from $1$ to $10$ (space components) as above.
\\
The commutators for the $O(10,1)$ gauge fields $k^{ab}$  are:
\begin{align}
    [k_{0}^{ab} (x), \ k_{0}^{cd} (y)] &= (\eta^{bc} {k_{0}}^{ad} - \eta^{da} {k_{0}}^{bc} - \eta^{bd} {k_{0}}^{ac} + \eta^{ca} {k_{o}}^{db}) \delta (\vec{x}- \vec{y}) \tag{60.1} \label{60.1} \\
    [k_{0}^{ab} (x), \ k_{M}^{cd} (y)] &= (\eta^{bc} {k_{M}}^{ad} - \eta^{da} {k_{M}}^{cb} - \eta^{bd} {k_{M}}^{ac} + \eta^{ca} {k_{M}}^{db}) \notag \\ 
    & \ \ \ \ \delta (\vec{x} - \vec{y}) + i C (\eta^{ca} \eta^{bd} - \eta^{cb} \eta^{ad}) \partial_{M} \delta (\vec{x} - \vec{y}) \tag{60.2} \label{eq:60.2} \\
    [k_{M}^{ab} (x), \ k_{N}^{cd} (y)] &= 0 \tag{60.3} \label{eq:60.3}.
\end{align}
The commutators that include ${B_{\mu}}^{abc} (x)$, are given by:
\begin{align}
    [B_{0}^{abc} (x), \ B_{\mu def} (y) ] &= -18 \ \delta_{[de}^{[ab}  k_{\mu f]}^{c]} (x) \delta (\vec{x} - \vec{y}) + 6 \ i C \delta_{def}^{abc} \partial_{\mu} \delta (\vec{x}-\vec{y}) \tag{61.1} \label{eq:61.1} \\
    [B_{M}^{abc} (x), \ B_{Ndef} (y)] &= 0 \tag{61.2} \label{eq:61.2} 
\end{align}
\begin{align}
    [{k_{\mu}}^{ab} &(x), \ {B_{0}}^{cde} (y)] = [{k_{0}}^{ab} (x), \ {B_{\mu}}^{ced} (y)] \notag \\
    &= \{ (\eta^{be} {B_{\mu}}^{ade} (x) + \eta^{bd} {B_{\mu}}^{aec} (x) + \eta^{be} {B_{\mu}}^{acd} (x)) - (a \longleftrightarrow b) \} \ \delta (\vec{x} - \vec{y}) \tag{61.3} \label{eq:61.3}
\end{align}
and
\begin{equation}
    [{k_{M}}^{ab} (x), \ {B_{N}}^{cde} (y)] = 0 \tag{61.4} \label{eq:61.4}.
\end{equation}
Finally, the (anti-)commutators, include the supersymmetry gauge field (current) or gravitino field, are:
\begin{align}
    &\ \ [k_{0}^{ab} (x), \ \psi_{\alpha, \mu} (y)] = [k_{\mu}^{ab} (x), \ \psi_{\alpha, 0} (y)] = {\chi_{\alpha}}^{\beta, ab} \psi_{\beta, \mu} \ \delta (\vec{x} - \vec{y}) \tag{62.1} \label{eq:62.1} \\
    &\ \ \{ \psi_{\alpha, 0} (x), \ \overline{\psi}_{\beta, 0} (y) \} \mid_{x_0 = y_0} = -2 \ e_{0}^{a} (x) \Gamma_{\alpha \beta}^{a} \ \delta (\vec{x} - \vec{y}) \tag{62.2} \label{eq:62.2} \\
    &\ \ \{ \psi_{\alpha, 0} (x), \ \overline{\psi}_{\beta, M} (y) \} \mid_{x_0 = y_0} \notag \\
    &\ \ \ \ \ \ \ \ \ \ \ \ \ = -2 \ e_{M}^{a} (x) \Gamma_{\alpha \beta}^{a} \ \delta (\vec{x} - \vec{y}) + i C \delta_{\alpha \beta} \partial_{M} \ \delta (\vec{x} - \vec{y}) \tag{62.3} \label{eq:62.3} \\
    &\ \ \{ \psi_{\alpha, 0} (x), \ \psi_{\beta, M} (y) \} \mid_{x_0 = y_0}  \notag \\
    &\ \ \ \ \ \ \ \ \ \ \ \ \ = -2 \ e_{M}^{a} (x) (\Gamma^{a} \Tilde{C}^{-1})_{\alpha \beta} \ \delta (\vec{x} - \vec{y}) + i \ C (\Tilde{C}^{-1})_{\alpha \beta} \partial_{M} \ \delta (\vec{x} - \vec{y}) \tag{62.4} \label{eq:62.4} \\
    &\ \ \{ \overline{\psi}_{\alpha, 0} (x), \ \overline{\psi}_{\beta, M} (y) \} \mid_{x_0 = y_0} \notag \\
    &\ \ \ \ \ \ \ \ \ \ \ \ \ = -2 \ e_{M}^{a} (x) (\Gamma^{a} \Tilde{C}_{\beta \alpha} \ \delta (\vec{x} - \vec{y}) + i \ C \Tilde{C}_{\beta \alpha} \partial_{M} \ \delta (\vec{x} - \vec{y}) \tag{62.5} \label{eq:62.5} \\
    &\ \ \{ \psi_{\alpha, M} (x), \ \overline{\psi}_{\beta, N} (y) \} \mid_{x_0 = y_0} = 0 \tag{62.6} \label{eq:62.6}.
\end{align}
$\Tilde{C}$ in the above equations is the charge conjugation operator, not to be confused with constant $C$.
\\
One caution is that all these commutators are for the anti-hermitian operators, since I omit ``$\,i\,$" in these commutators in constant to equation (\ref{eq:2}) or (\ref{eq:3}).
\\
\\
The equations of motion are derived using the above local commutators together with equations (\ref{eq:8}), (\ref{eq:9}), (\ref{eq:10}) and (\ref{eq:11}).  
I list the result of calculations:
\\
(1) Equations for the internal energy- momentum gauge field/current or the elfbein field $e_{\mu}^{a} (x)$:
\begin{equation}
    D_{\mu c}^{a} e_{\nu}^{c} - D_{\nu c}^{a} e_{\mu}^{c} = 0 \tag{63.1} \label{eq:63.1}
\end{equation}
with 
\begin{equation}
    D_{\mu c}^{a} = \delta_{c}^{a} \partial_{\mu} + \frac{i}{2C} k_{\mu c}^{a} (x) \tag{63.2} \label{eq:63.2}.
\end{equation}
I also have a consistent condition that:
\begin{equation}
    k_{\mu}^{bc} (x) (\eta^{ba} e_{\nu}^{c} (y) - \eta^{ca} e_{\nu}^{b} (y) \mbox{ must be antisymmetric in } \mu \mbox{ and } \nu \tag{63.3} \label{eq:63.3}.
\end{equation}
Lastly, I get the conservation law:
\begin{equation}
    \partial_{0} e_{0}^{a} - \partial_{M} e_{M}^{a} = 0 \tag{63.4} \label{eq:63.4}.
\end{equation}
(2) Equation for the $O(10,1)$ gauge field/current or the connection field ${k_{\mu}}^{ab} (x)$:
\begin{align}
    &{D^{cd, ab}}_\mu k_{\nu ab} (x) - {D^{cd, ab}}_\nu k_{\mu ab} (x) \notag \\
    &= \frac{i}{c} ({e^c}_\mu (x) e_\nu^d (x) - {e^d}_\mu (x) e_\nu^c (x)) - \frac{i}{2c} \{ \eta^{bc} {B_\nu}^{ade} - (a \longleftrightarrow b) \} B_{\mu abe} (x) \notag \\
    &- \frac{i}{2c} (\chi_{\alpha}^{\beta, cd} \overline{\psi}_{\beta, \nu} \psi_{\alpha, \mu} (y) - (\mu \longleftrightarrow \nu)) \tag{64.1} \label{eq:64.1}
\end{align}
with
\begin{equation}
    {D^{cd, ab}}_{\mu} = \eta^{ca} \eta^{db} \partial_{\mu} - i / 2 (\eta^{bc} {k_{\mu}}^{ad} - \eta^{da} {k_\mu}^{cd}) \tag{64.2} \label{eq:64.2}.
\end{equation}
Consistent conditions are:
\begin{equation}
    (\eta^{bc} {k_{\mu}}^{ad} - \eta^{da} {k_{\mu}}^{cb} - \eta^{bd} {k_{\mu}}^{ac} + \eta^{ca} {k_{\mu}}^{db}) k_{\nu cd} \mbox{ must be antisymmetric in } \mu, \nu.  \tag{64.3} \label{eq:64.3}
\end{equation}
\begin{equation}
    \{ \eta^{bc} {B_{\nu}}^{ade} - (a \longleftrightarrow b) \} B_{\mu cde} (x) \mbox{ must be antisymmetric in } \mu \longleftrightarrow \nu. \tag{64.4} \label{eq:64.4}
\end{equation}
\begin{equation}
    \chi_{\alpha}^{\beta, ab} \overline{\psi}_{\beta, \gamma} \psi_{\alpha, \mu} (y) \mbox{ must be anti-symmetric in } \mu \longleftrightarrow \nu \tag{64.5} \label{eq:64.5}.
\end{equation}
I also get the conservation law:
\begin{equation}
    \partial_{0} k_{0}^{ab} - \partial_{M} k_{M}^{ab} = 0 \tag{64.6} \label{eq:64.6}.
\end{equation}
(3) Equation for the antisymmetric gauge field/current $B_{\mu}^{abc} (x)$:
\begin{align}
    \partial_{\mu} B_{\nu}^{abf} & (x)  - \partial_{\nu} B_{\mu}^{abf} (x) \notag \\
    & = \frac{i}{C} (k_{\mu e}^{[f} \ B_{\nu}^{ab] e} - (\mu \longleftrightarrow \nu)) + \frac{i}{C} {\xi^{\beta, abf}}_{\alpha} \overline{\psi}^{\alpha}_{\mu} (x) \psi_{\beta, \nu} (x) \tag{65.1} \label{eq:65.1}
\end{align}
with the constraint condition
\begin{equation}
    {\xi^{\beta}}_{\alpha, \ abc} \overline{\psi}^{\alpha}_{\mu} (x) \psi_{\beta, \nu} (x) \mbox{ must be antisymmetric in } \mu \longleftrightarrow \nu \tag{65.2} \label{eq:65.2}.
\end{equation}
I have the conservation law
\begin{equation}
    \partial_{0} B_{0}^{abf} (x) - \partial_{M} B_{M}^{abf} (x) = 0 \tag{65.3} \label{eq:65.3}.
\end{equation}
(4) Finally, I get the equation for the supersymmetry gauge field/current or the gravitino field:
\begin{align}
    \partial_{\mu} \psi_{\alpha \nu} (x) &- \partial_{\nu} \psi_{\alpha \mu} (x) = \frac{i}{C} (2 e_{\nu}^{a} (x) \Gamma_{\alpha \beta}^{a} \psi_{\beta, \mu} (x) \notag \\
    &+ \frac{1}{2} {\chi_{\alpha}}^{\beta, cd} k_{\mu cd} (x) \psi_{\beta, \nu} (x)) + \frac{i}{6C} {\zeta^{\beta, cde}}_{\alpha} \psi_{\beta, \nu} (x) B_{\mu cde} (x) \tag{66.1} \label{eq:66.1}.
\end{align}
The consistent conditions are:
\begin{equation}
    2 e_{\nu}^{a} (x) \Gamma_{\alpha \beta}^{a} \psi_{\beta, \mu} (x) + \frac{1}{2} {\chi_{\alpha}}^{\beta, cd} k_{\mu cd} (x) \psi_{\beta, \nu} (x) \mbox{ must be antisymmetric in } \mu \longleftrightarrow \nu \tag{66.2} \label{eq:66.2}
\end{equation}
\begin{equation}
    {\zeta^{\beta, cde}}_{\alpha} \psi_{\beta, \nu} (x) B_{\mu cde} (x) \mbox{ must be antisymmetric in } \mu \longleftrightarrow \nu \tag{66.3} \label{eq:66.3}.
\end{equation}
I also have the conservation law:
\begin{equation}
    \partial_{0} \psi_{\alpha 0} (x) - \partial_{M} \psi_{\alpha, M} (y) = 0 \tag{66.4} \label{eq:66.4}.
\end{equation}
The conservation law guarantees that the theory is locally supersymmetric.  
\\
The completely new feature of our theory is that the supersymmetry generator is given by
\begin{equation}
    \Psi^{\alpha} = \int {\Psi^{\alpha}}_{0} (x) \mathrm{d} \vec{x}  \tag{67} \label{eq:67},
\end{equation}
that is, the space integral of the time component of the gravitino field.  
The tangent space energy-momentum is given by
\begin{equation}
    P_{a} = \int e_{0a} (x) \mathrm{d} \vec{x} \tag{68} \label{eq:68},
\end{equation}
and the supersymmetry algebra is
\begin{equation*}
    \{ \Psi^{\alpha}, \overline{\Psi}^{\beta} \} = -2 P_{a} \Gamma^{\alpha \beta, a} + \xi^{\alpha \beta, abc} S_{abc} + \dots
\end{equation*}
as is given in equation (\ref{eq:56}).

\section{Discussion of the gravity equations}

Here, I discuss the implications of the equations for the elfbein field $e_{\mu}^{a} (x)$ and the connection field $k_{\mu}^{ab} (x)$.
First, from equations (\ref{eq:63.1}) and (\ref{eq:63.2}), $k_{\mu}^{ab} (x)$ is related to the usual connection field $\omega_{\mu}^{ab} (x)$ as:
\begin{equation}
    \omega_{\mu c}^{a} = \frac{i}{2C} k_{\mu c}^{a} (x) \tag{69} \label{eq:69}.
\end{equation}
Then equation (\ref{eq:63.1}) is the usual relation between the elfbein and the connection field, thus providing the usual expression of the latter in terms of the former.
\\
The next question is the general covariance of the equations. 
In deriving these equations, I assumed that I could take a locally flat system at any point in the curved space-time.  
If I can write these equations in a generally covariant way, it would show that the equations are valid at any point in the curved space-time.
\\
I assume that the background metric is not completely arbitrary but satisfies the equations derived above, including the conservation equation for the $e_{\mu}^{a} (x)$.  
I denote the background field variables such as the elfbein field by capping them with ``$ \; \Tilde \;$.''
\begin{equation}
    \Tilde{g}_{\mu\nu} (x) = \Tilde{e}_{\mu}^{a} (x) \Tilde{e}_{\nu}^{a} (x) \tag{70} \label{eq:70}.
\end{equation}
The summation over ``$a$" must be understood to be in the Lorentzian tangent space. 
\\
The affine connection $\Gamma_{\mu\lambda}^{\nu} (x)$ is defined as usual:
\\
I define
\begin{equation}
    \Gamma_{\mu\lambda}^{\nu} = e_{a}^{\nu} (\partial_{\mu} e_{\lambda}^{a} + e_{\lambda}^{b} \omega_{\mu b}^{a}) \tag{71} \label{eq:71}.
\end{equation}
Then,
\begin{align}
    \Gamma_{\mu \lambda}^{\nu} + \Gamma_{\lambda \mu}^{\nu} &= e_{a}^{\nu} (\partial_{\mu} e_{\lambda}^{a} + \partial_{\lambda} e_{\mu}^{a}), \tag{72.1} \label{eq:72.1} \\
    \Gamma_{\mu \lambda}^{\nu} - \Gamma_{\lambda \mu}^{\nu} &= e_{a}^{\nu} (\partial_{\mu} e_{\lambda}^{a} + e{\lambda}^{b} \omega_{\mu b}^{a} - (\mu \longleftrightarrow \lambda)) = 0, \tag{72.2} \label{eq:72.2}
\end{align}
because
\begin{equation*}
    \partial_{\mu} e_{\nu}^{a} - \partial_{\nu} e_{\mu}^{a} = - \omega_{\mu c}^{a} (x) e_{\nu}^{c} (y) + \omega_{\nu c}^{a} (x) e_{\mu}^{c} (y),
\end{equation*}
and $e_{\lambda}^{b} \omega_{\mu b}^{a}$ is antisymmetric in $\mu \longleftrightarrow \lambda$ as is obtained in equation (\ref{eq:63.3}).
\\
Equations (\ref{eq:71}), (\ref{eq:72.1}) and (\ref{eq:72.2}) must be true for the background metric by an assumption.  
The symmetric nature of $\Gamma_{\mu \lambda}^{\nu}$ under $\mu$, $\lambda$ is enough to prove that for any tensor $A_{\mu}^{abc\dots} (x)$:
$B_{\mu \nu}^{abc \dots} (x) = \partial_{\mu} A_{\nu}^{abc \dots} (x)  - \partial_{\nu} A_{\mu}^{abc \dots} (x)$ is a generally covariant antisymmetric tensor:
\begin{equation}
    B_{\mu \nu}^{\prime \ abc \ \dots} (x') = \frac{\partial x^{\rho}}{\partial x^{\prime \mu}} \frac{\partial x^{\kappa}}{\partial x^{\prime \nu}} B_{\rho \kappa}^{abc \dots} (x) \tag{73} \label{eq:73}.
\end{equation}
All the equations which contain antisymmetric derivatives are, therefore, generally covariant.  
I also have
\begin{equation}
    \Gamma_{\mu \lambda}^{\nu} = \frac{1}{2} e_{a}^{\nu} (\partial_{\mu} e_{\lambda}^{a} + \partial_{\lambda} e_{\mu}^{a}), \ \ \eta^{\mu \lambda} \Gamma_{\mu \lambda}^{\nu} = 0 \tag{74} \label{eq.74}
\end{equation}
\begin{equation}
    D_{\mu} A^{\mu} = D_{\mu} g^{\mu \rho} A_{\rho} = g^{\mu \rho} D_{\mu} A_{\rho} \longrightarrow \eta^{\mu \rho} D_{\mu} A_{\rho} = (\eta^{\mu \nu} \partial_{\mu} - \eta^{\mu \rho} \Gamma_{\mu \rho}) A_{\nu} = 0 \tag{75} \label{eq:75}
\end{equation}
where ``$\longrightarrow$'' means going to the flat metric at this point.
This is because
\begin{equation}
  (\delta_{\rho}^{\mu} \partial_{\mu} + \Gamma_{\mu \rho}^{\mu}) A^{\rho} = (\partial_{\mu} g^{\mu \rho} + \Gamma_{\mu \nu}^{\mu} g^{\nu \rho} + \Gamma_{\mu \nu}^{\rho} g^{\mu \nu}) A_{\rho} + g^{\mu \rho} (\delta_{\rho}^{\nu} \partial_{\mu} - \Gamma_{\mu \rho}^{\nu}) A_{\nu} \tag{76} \label{eq:76}.
\end{equation}
Here, I used
\begin{equation}
    D_\mu g^{\mu \rho} = \partial_\mu g^{\mu \rho} + \Gamma_{\mu \nu}^\mu g^{\nu \rho} + \Gamma_{\mu \nu}^\rho g^{\mu \nu} = 0 \tag{77} \label{eq:77}.
\end{equation}
This equation (\ref{eq:75}) holds for $A_{\mu}^{(abc \dots)}$, because
\begin{gather}
    (\delta_\rho^\mu \partial_\mu + \Gamma_{\mu \rho}^\mu + \Omega_\rho) A^{\rho, abc \dots)} = (\delta_\rho^\mu \partial_\mu + \Gamma_{\mu \rho}^\mu) A^{\rho, abc \dots)} \tag{78} \label{eq:78} \\
    \Omega_\rho, \ A^{\rho, abc \dots)} = g^{\rho \mu} \Omega_\rho, \ A_\mu^{(abc \dots)} = 0 \tag{79} \label{eq:79}.
\end{gather}
This concludes the proof that the conservation law is also generally covariant.  
The next question is to examine the equations in connection with the usual Einstein-Hilbert theory.  It is still not clear how the latter is quantized. 
I start with  
\begin{align}
    {D^{cd, ab}}_\mu k_{\nu ab} (x) &- {D^{cd, ab}}_\nu k_{\mu ab} (x) \notag \\
    &= \frac{i}{c} ({e^c}_\mu (x) e_\nu^d (x) - {e^c}_\nu (x) e_\mu^d (x)) \notag \\
    &- \frac{i}{2c} \{ \eta^{bc} {B_\nu}^{ade} - (a \longleftrightarrow b) \} B_{\mu cde} (x) - (\mu \longleftrightarrow \nu) \} \notag \\
    &- \frac{i}{2c} ({\chi_\alpha}^{\beta, ab} \overline{\psi}_{\beta, \gamma} \psi_{\alpha, \ \mu} (y) - (\mu \longleftrightarrow \nu)) \tag{64.1}
\end{align}
with
\begin{equation*}
    {D^{cd, \ ab}}_\mu = \eta^{ca} \eta^{db} \partial_\mu - i / 2 \ C (\eta^{bc} {k_\mu}^{ad} - \eta^{da} {k_\mu}^{cb}) = \eta^{ca} \eta^{db} \partial_\mu - (\eta^{bc} {\omega_\mu}^{ad} - \eta^{da} {\omega_\mu}^{cb}).
\end{equation*}
Using equation (\ref{eq:69}), I have
\begin{align}
    {D^{cd, ab}}_\mu & \ \omega_{\nu ab} (x) - {D^{cd, ab}}_\nu \ \omega_{\mu ab} (x) \notag \\
    &= \frac{-1}{2} ({e^c}_\mu (x) \ e_\nu^d (x) - {e^c}_\nu (x) \ e_\mu^d (x)) \notag \\
    &- \frac{1}{4} \{ \eta^{bc} {B_\nu}^{abe} - (a \longleftrightarrow b) B_{\mu cde} (x) - (\mu \longleftrightarrow \nu) \} \notag \\
    &+ \frac{1}{2} \{ {\chi_\alpha}^{\beta, ab} \overline{\psi}_{\beta, \gamma} \psi_{\alpha, \mu} (y) - (\mu \longleftrightarrow \nu) \} \tag{80} \label{eq:80}.
 \end{align}
Then by using the definition of the curvature tensor in terms of the spin tensor:
\begin{equation}
    R_{\sigma \mu \nu}^{\lambda} = e_c^{\lambda} \ e_\sigma^d (\partial_{\mu} \ {\omega^c}_{\nu d} (x) + \omega_{\mu b}^c \ {\omega^b}_{d \nu} (x) - (\mu \longleftrightarrow \nu) \tag{81} \label{eq:81}.
\end{equation}
I get
\begin{align}
    R_{\sigma \mu \nu}^{\lambda} &= e_c^\lambda \ e_\sigma^d (\partial_\mu \ {\omega^c}_{\nu d} (x) + {\omega_{\mu b}}^c \ {\omega^b}_{d \nu} (x) - (\mu \longleftrightarrow \nu)) \notag \\
    &= \frac{1}{2} ({\delta^\lambda}_\mu \ g_{\nu \sigma} (x) - {\delta^\mu}_{\nu} \ g_{\mu \sigma} (x)) + \{ e^{b \lambda} \ e_\sigma^d \ \omega_{\mu ad} \ {\omega^a}_{\nu b} (x) - (\mu \longleftrightarrow \nu) \} \notag \\
    &+ \frac{1}{4} \{ (e_c^\lambda \ e_\sigma^a - e^{a \lambda} \ e_{c \sigma}) {B_\nu}^{cbe} B_{\mu abe} (x) - (\mu \longleftrightarrow \nu) \} \notag \\
    &- \frac{1}{2} e_c^\lambda \ e_{d \sigma} ({\chi_\alpha}^{\beta, cd} \overline{\psi}_{\beta, \nu} \psi_{\alpha, \mu} (x) - (\mu \longleftrightarrow \nu)) \tag{82} \label{eq.82}.
\end{align}
Here, I used the definition of $g_{\mu \nu} (x)$ in terms of elfbein and also changed all the annihilation operators to hermitian operators by:
\begin{equation}
    e, \ B \ \mbox{and } \Psi \longrightarrow i \ (e, \ B \mbox{ and } \psi) \tag{83} \label{eq:83}.
\end{equation}
Finally I get,
\begin{align}
    R_{\sigma^\nu} &= R_{\sigma \mu \nu}^{\mu} \notag \\
    &= \frac{D-1}{2} g_{\nu \sigma} (x) + e^{b \mu} \ e_\sigma^d \{ \omega_{\mu ad} \ {\omega^a}_{\nu b} (x) - \omega_{\nu ad} \ {\omega^a}_{\mu b} (x) \notag \\
    &+ \frac{1}{4} \{ e_c^\mu \ e_\sigma^a - e^{a \mu} \ e_{c \sigma} \} \{ {B_\nu}^{cbe} B_{\mu abe} (x) - (\mu \longleftrightarrow \nu) \} \notag \\
    &- \frac{1}{2} e_c^\mu \ e_{d \sigma} \{ {\chi_\alpha}^{\beta, cd} (\overline{\psi}_{\beta, \nu} \ \psi_{\alpha, \mu} (x) - (\mu \longleftrightarrow \nu)) \} \tag{84} \label{eq:84}.
\end{align}
The first term on the right hand side is the cosmological constant (anti de Sitter) which originates from the contribution of elfbein field $e_{\mu}^{a} (x)$.  
The second term is a kind of back reaction (gravitational energy contribution to the curvature).  
The third and the fourth terms are contributions from the bosonic gauge and supersymmetry gauge fields.  
The equation is not exactly that of Einstein-Hilbert. 
\\
What it gives me when I take the classical limit of equation (\ref{eq:84}) is a major concern and it will be investigated in future work.

\section{Physical constants}

I now discuss the physical constants which appear in this theory.  
The space-time coordinate $x_{\mu}$ can be treated as dimensionless by measuring in terms of fundamental length constant $l$.  
All the currents/fields are assumed to be dimensionless.  
Then the commutation relations do not have any constants appearing explicitly.  
\\
\\
The energy momentum tensor  can be written as
\begin{equation}
    \frac{\hbar}{l} \int \mathrm{d}x \ \Theta_{MN} (x) = \frac{\hbar}{l} \int \mathrm{d}x [tr (\Omega_M \Omega_N) - \frac{1}{2} \eta_{MN} tr (\Omega_L \Omega_L)] \tag{85} \label{eq:85}.
\end{equation}
Then the equation of motion is
\begin{equation}
    \frac{\hbar}{l} i \partial_{x} f = [f, \ \frac{\hbar}{l} \int \mathrm{d}x (tr (\Omega_M \Omega_N) - \frac{1}{2} \eta_{MN} tr (\Omega_L \Omega_L))] \tag{86} \label{eq:86}.
\end{equation}
Therefore, the factor $\frac{\hbar}{1}$ disappears from the equation of motion completely.  
When I recover the space-time coordinate with the dimension of length, I recover $C$, with 
\begin{equation}
    C = l^{11} \tag{87} \label{eq:87}.
\end{equation}
Then the delta function commutator such as 
\begin{equation*}
    [e_\nu^c (y) k_0^{ab} (x)] \mid_{x_0=Y_0} = (\eta^{ac} e_\nu^b (y) - \eta^{bc} e_\nu^a (y)) \delta (\vec{x} - \vec{y}) 
\end{equation*}
must be modified to 
\begin{align}
    [e_\nu^c (y) k_0^{ab} (x)] \mid_{x_0=y_9} &= (\eta^{ac} e_\nu^b (y) - \eta^{bc} e_\nu^a (y)) \ l^{10} \delta (\vec{x} - \vec{y}) \notag \\
    &= (\eta^{ac} e_\nu^b (y) - \eta^{bc} e_\nu^a (y)) ^{10/11} \delta (\vec{x} - \vec{y}) \notag, 
\end{align}
and
\begin{equation*}
    \frac{\hbar}{l} \int \mathrm{d}x \ [f_M^a (x) f_N^a (x) - \frac{1}{2} \eta_{MN} f_L^a (x) f_L^a (x)]
\end{equation*}
must be changed to
\begin{equation*}
    \frac{\hbar}{C} \int \mathrm{d}x \ [f_M^a (x) f_N^a (x) - \frac{1}{2} \eta_{MN} f_L^a (x) f_L^a (x)].
\end{equation*}
Einstein equation with physical constants becomes
\begin{align}
    R_{\sigma \nu} &= R_{\sigma \mu \nu}^\mu \notag \\
    &= \frac{D-1}{2 l} g_{\nu \sigma} (x) + \frac{1}{l} \{ e^{b \mu} \ e_\sigma^d (\omega_{\mu ad} \ {\omega^a}_{\nu b} (x) - \omega_{\nu ad} \ {\omega^a}_{\mu b} (x)) \notag \\
    &+ \frac{1}{4} (e_c^\mu \ e_\sigma^a - e^{a \mu} \ e_{c \sigma}) ({B_\nu}^{cbe} B_{\mu abe} (x) - (\mu \longleftrightarrow \nu)) \notag \\
    &- \frac{1}{2} e_c^\mu \ e_{d \sigma} ({\chi_\alpha}^{\beta, cd} \overline{\psi}_{\beta, \nu} \ \psi_{\alpha, \mu} (x) - (\mu \longleftrightarrow \nu)) \} \tag{88} \label{eq:88}.
\end{align}
Define 
\begin{align}
    S_{\sigma \nu} &= \frac{\hbar}{l^D} \{ e^{b \mu} \ e_\sigma^d (\omega_{\mu ad} \ {\omega^a}_{\nu b} (x) - \omega_{\nu ad} \ {\omega^a}_{\mu b} (x)) \notag \\
    &+ \frac{1}{4} (e_c^{\mu} \ e_\sigma^a - e^{a \mu} \ e_{c \sigma}) ({B_\nu}^{cbe} B_{\mu abe} (x) - (\mu \longleftrightarrow \nu)) \notag \\
    &- \frac{1}{2} e_c^\mu \ e_{d \sigma} ({\chi_\alpha}^{\beta, cd} \overline{\psi}_{\beta, \nu} \ \psi_{\alpha, \mu} (x) - (\mu \longleftrightarrow \nu)) \} \tag{89} \label{eq:89}.
\end{align}
I get
\begin{equation}
    R_{\sigma \nu} = R_{\sigma \nu \nu}^{\mu} = \frac{D-1}{l} g_{\nu \sigma} (x) + \frac{C}{\hbar l} S_{\sigma \nu} = \frac{D-1}{l} g_{\nu \sigma} (x) + \frac{l^{D-1}}{\hbar} S_{\sigma \nu} \tag{90} \label{eq:90}.
\end{equation}
$\omega_{\mu ad}$ has the dimension of $1/l$ by definition.  
Therefore, the above equation becomes
\begin{equation}
    R_{\sigma \nu}  = R_{\sigma \mu \nu}^{\mu} = \frac{D-1}{l^2} g_{\nu \sigma} (x) + \frac{C}{\hbar l^2} S_{\sigma \nu} = \frac{D-1}{l^2} g_{\nu \sigma} (x) + \frac{l^{D-2}}{\hbar} S_{\sigma \nu} \tag{91} \label{eq:91}.
\end{equation}
Suppose this was the case of 4-dimensional space-time, I get
\begin{equation}
    R_{\sigma \nu} = \frac{3}{l^2} g_{\nu \sigma} (x) + \frac{C}{\hbar l^2} S_{\sigma \nu} = \frac{3}{l^2} g_{\nu \sigma} (x) + \frac{l^2}{\hbar} S_{\sigma \nu} \tag{92} \label{eq:92}.
\end{equation}
Suppose I use the Planck length for $l: l=\sqrt{G \hbar}$, I get
\begin{equation}
    R_{\sigma \nu} - \frac{3}{\hbar G} \ g_{\nu \sigma} (x) = GS_{\sigma \nu} \tag{93} \label{eq:93}.
\end{equation}
$S_{\sigma \nu}$ can be written as
\begin{align}
    S_{\sigma \nu} &= \frac{\hbar}{l^D} \{ e^{b \mu} \ e_\sigma^d (\omega_{\mu ad} \ {\omega^a}_{\nu b} (x) - \omega_{\nu ad} \ {\omega^a}_{\mu b} (x) \} \notag \\
    & \ \ \ \ + \frac{1}{4} (e_c^u \ e_\sigma^a - e^{a \mu} \ e_{c \sigma}) ({B_\nu}^{cbe} B_{\mu abe} (x) - (\mu \longleftrightarrow \nu)) \notag \\
    & \ \ \ \ - \frac{1}{2} e_c^{\mu} \ e_{d \sigma} ({\chi_\alpha}^{\beta, cd} \overline{\psi}_{\beta, \nu} \ \psi_{\alpha, \mu} (x) - (\mu \longleftrightarrow \nu)) \} \notag \\
    &= \frac{\hbar}{l^D} \{ (e^{b \mu} \ e_\sigma^d - e^{d \mu} \ e_\sigma^b) \ \omega_{\mu ad} \ {\omega^a}_{\nu b} (x) \notag \\
    & \ \ \ \ - \frac{1}{2} (-e^{d \mu} e_{b \sigma} + e_b^{\mu} e_{\sigma}^d) B_{\mu dfe} (x) {B_\nu}^{bfe} \notag \\
    & \ \ \ \ - \frac{1}{2} (e_b^\mu e_{d \sigma} - e_d^\mu e_{b \sigma}) ({\chi_\alpha}^{\beta, bd} \overline{\psi}_{\beta, \nu} \psi_{\alpha, \mu} (x)) \} \tag{94} \label{eq:94}.
\end{align}
Proof is as follows:
\begin{align}
    {\chi_\alpha}^{\beta, bd} \overline{\psi}_{\beta, \nu} \ \psi_{\alpha, \mu} (x) &= - {\chi_\alpha}^{\beta, bd} C_{\gamma \beta} \ \psi_{\alpha, \mu} (x) \ \psi_{\gamma, \nu} \notag \\
    &= {\chi_\alpha}^{\beta, bd} C_{\gamma \beta} C_{\alpha \epsilon} \ \psi_{\delta, \mu} (x) \ \psi_{\gamma, \nu} \notag \\
    &= - C_{\delta \alpha} \ {\chi_\alpha}^{\beta, bd} C_{\gamma \beta} \ \overline{\psi}_{\delta, \mu} (x) \ \psi_{\gamma, \nu} \notag \\
    &= - {\chi_\delta}^{\gamma, bd} \ \overline{\psi}_{\delta, \mu} (x) \ \psi_{\gamma, \nu} \tag{95} \label{eq:95}
\end{align}
with 
\begin{equation}
    [J_{ab}, \ \Psi^\alpha C_{\alpha \gamma}] = - C_{\beta \epsilon} \ {\chi^\alpha}_{\beta, ab} C_{\alpha \gamma} C_{\delta \epsilon} \ \Psi^\epsilon \tag{96} \label{eq:96}
\end{equation}
and 
\begin{equation}
    C_{\beta \epsilon} \ {\chi^\alpha}_{\beta, ab} C_{\alpha \gamma} = - {\chi^\gamma}_{\epsilon, ab} \tag{97} \label{eq:97}.
\end{equation}
Finally, 
\begin{equation}
    S_{\sigma \nu} = \frac{\hbar}{l^D} (e^{b \mu} e_\sigma^d - e^{d \mu} e_\sigma^b) \{ \omega_{\mu ad} \ {\omega^a}_{\nu b} (x) - \frac{1}{2} \eta_{b \overline{b}} B_{\mu dfe} (x) {B_\nu}^{\overline{b} fe} - \frac{1}{2} \eta_{b \overline{b}} \eta_{d \overline{d}} ({\chi_\alpha}^{\beta, \overline{b} \overline{d}} \overline{\psi}_{\beta, \nu} \ \psi_{\alpha, \mu} (x)) \} \tag{98} \label{eq:98}.
\end{equation}
This is not exactly the energy-momentum tensor $\Theta_{\mu \nu} (x)$ which I defined in equation (\ref{eq:8}) as the bilinear of the current/field.  
The equivalence principle of the gravity is in this sense violated in the quantum level in my formulation.  
What it means in the classical limit remains to be seen.

\section{Nonlinear realization}

No important role is played by the nonlinear realization of the currents/fields in the formulation of M-theory using the current algebra as explained in the above sections.  
But I can write all the currents/fields in terms of curved space-time scalars and possibly write down Lagrangian for these fields although I emphasize that it is not at all needed.  
The fact that the nonlinear realization of the currents/fields leads to a $\sigma$-model Lagrangian ($tr \int \partial_{\mu} g^{\-1} \partial^{\mu} g \ \mathrm{d} x$) was pointed out long time ago by M. Yoshimura and the present author\cite{HSandMY}.
\\
I start with
\begin{align}
    \Omega_\mu (x) &= g^{-1} \partial_\mu \ g = \exp (- \phi^a T_a) \ \partial_\mu \exp (\phi^a T_a) \notag \\
    &= \partial_\mu \ \phi^a T_a + \frac{1}{2} \ (\partial_\mu \ \phi^a) \ \phi^b [T_a, \ T_b] + \frac{1}{6} \ (\partial_\mu \ \phi^a) \ \phi^b \ \phi^c [[T_a, \ T_b], T_c] + \dots \tag{99} \label{eq:99}.
\end{align}
I apply this formula to the following currents/fields:
\begin{align}
    {f_\mu}^a (x) \ T_a &= {k_{\mu a}}^b (x) K_b^a + {B_\mu}^{abc} (x) R_{abc} + \dots \tag{100} \label{eq:100} \\
    w_\mu^a H_a &= e_\mu^a P_a + \overline{\psi}_{\alpha, \mu} (x) \ \Psi_a \dots \tag{101} \label{eq:101} \\
    g^{-1} \partial_{\mu} \ g &= {f_\mu}^a (x) \ T_a + w_\mu^a H_a \notag \\
                                     &= \exp (-\phi^a T_a + \Tilde{\phi} H_a) \ \partial_\mu \exp (\phi^a T_a + \Tilde{\phi} H_a) \tag{102} \label{eq:102}.
\end{align}
The results are the following:
\begin{align}
    &k_{\mu ab} (x) = \partial_\mu \ k_{ab} (x) + \frac{1}{2} (\eta^{fc} \partial_\mu \ k_{af} \ k_{cb} - \eta^{de} \partial_\mu \ k_{eb} \ k_{ad} \notag \\
    & \ \ \ \ \ \ \ \ \ \ \ \ \ \ \ \ \ \ \ \ \ \ \ \ \ \ \ \ \ \ \ \ \ \ \ \ \ \ \ \ \ \ \ \ \  \ \ \ \ \ \ - \eta^{fd} \partial_\mu \ k_{af} \ k_{bd} + \eta^{ce} \partial_\mu \ k_{eb} \ k_{ca}) + \dots \tag{103} \label{eq:103} \\
    &e_\mu^a = \partial_{\mu} \ \phi^a + \eta_{bc} (\partial_\mu k^{ab} \phi^c - k^{ab} \partial_\mu \ \phi^c) - (\partial_\mu \ \overline{\psi}^\alpha \Gamma^{\alpha \beta, a} \ \psi^\beta - \overline{\psi}^\alpha \Gamma^{\alpha \beta, a} \partial_\mu \ \psi^\beta)  + \dots \tag{104} \label{eq:104} \\
    &B_{\mu abc} (x) = \partial_{\mu} \ A_{abc} + \eta^{gh} \partial_\mu \ k_{ag} \ A_{hbc} + \eta^{gd} \partial_\mu \ k_{ag} \ A_{cdb} + \eta^{ge} \partial_\mu \ k_{ag} \ A_{bce} + \dots \tag{105} \label{eq:105} \\
    &{\psi_\mu}^\alpha = \partial_\mu \ \psi^\alpha + \frac{1}{2} \ {\chi^\alpha}_{\beta, ab} (\partial_\mu \ k^{ab} \psi^\beta - k^{ab} \partial_\mu \ \psi^\beta) \notag \\
    & \ \ \ \ \ \ \ \ \ \ \ \ \ \ \ \ \ \ \ \ \ \ \ \ \ \ \ \ \ \ \ \ \ \ \ \ \ \ \ \ \ \ \ \ \ \ \ + \frac{1}{2} {\zeta^\alpha}_{\beta, abc} (\partial_\mu \ A^{abc} \ \psi^\beta - A^{abc} \partial_\mu \ \psi^\beta) + \dots \tag{106} \label{eq:106}.
\end{align}

\section*{Acknowledgment}

This work was  inspired by a talk by Professor P. West presented at the Abdus Salam Memorial Symposium held in Singapore in January, 2016.  
I deeply appreciate him for his presentation and conversation afterwards.  
\\
\\
I also want to express my sincere thanks to Professor L. Brink for persuading me to attend this symposium.  
Major part of this work was done when the author was visiting UCLA, summer 2016.  
I want to thank Professors R. Peccei and A. Kusenko for their hospitality.  

\section*{Appendix: Calculation of $\chi^{\beta}_{\alpha, ab}$}

$\chi^{\beta}_{\alpha, ab}$ is defined by 
\begin{equation}
    [J_{ab}, \ \Psi^{\alpha}] = {\chi^{\alpha}}_{\beta, ab} \Psi^{\beta} \tag{A1} \label{eq:A1}.
\end{equation}
Here,
\begin{equation}
    J_{ab} = \frac{-i}{4} \ [\Gamma_{a}, \ \Gamma_{b}] \tag{A2} \label{eq:A2}
\end{equation}
where $\Gamma$ are Dirac Matrices.  
\\
I start with creation and annihilation operators
\begin{equation}
    \{ \Gamma_{a+}, \ \Gamma_{b-} \} = \delta_{ab}, \ \ \{ \Gamma_{a+}, \ \Gamma_{b+} \} = \{ \Gamma_{a-}, \ \Gamma_{b-} \} = 0, \mbox{ for } a,b= 0, 1, 2, 3, 4 \tag{A3} \label{A3},
\end{equation}
then
\begin{align}
        &\Gamma_0 = \Gamma_{0+} + \Gamma_{0-} \notag \\
        &\Gamma_1 = \Gamma_{0+} + \Gamma_{0-} \notag \\
        &\Gamma_{2a} = \Gamma_{a+} + \Gamma_{a-} \notag \\
        &\Gamma_{2a+1} = i \ (\Gamma_{a+} - \Gamma_{a-}) \mbox{ for } a=1, 2, 3, 4 \notag \\
        &\Gamma_{10} = \Gamma_0 \Gamma_1 \dots \Gamma_9 \tag{A4} \label{eq:A4}.
\end{align}
Spinor representation is given by
\begin{equation*}
    \Psi_\alpha = (\Gamma_{0+})^{\alpha_{0} + 1/2} (\Gamma_{1+})^{\alpha_{1} + 1/2} \dots (\Gamma_{4+})^{\alpha_{4} + 1/2}, \ \ \alpha_a = \pm \frac{1}{2}, \ \ \alpha = \{ \alpha_0, \alpha_1, \alpha_2, \alpha_3, \alpha_4 \}.
\end{equation*}
Define
\begin{align}
    J_{a+, \ b+} = = \frac{-i}{2} \Gamma_{a+} \Gamma_{b+}, &\ J_{a-, \ b-} = = \frac{-1}{2} \Gamma_{a-} \Gamma_{b-} \notag \\
    J_{a+, \ b-} = = \frac{-1}{2} (\Gamma_{a+} \Gamma_{b-} - \frac{1}{2} \delta_{ab}), &\ J_{a-, \ b+} = = \frac{-i}{2} (\Gamma_{a-} \Gamma_{b+} - \frac{1}{2} \delta_{ab}) \tag{A5} \label{eq:A5},
\end{align}
then
\begin{equation}
    [J_{a \pm, b \pm}, \ \Psi_\alpha] = [\Gamma_{a \pm} \Gamma_{b \pm}, \ \Psi_\alpha] = \Gamma_{a \pm} \{ \Gamma_{b \pm}, \ \Psi_\alpha \} - \{ \Gamma_{a \pm}, \ \Psi_\alpha \} \Gamma_{b \pm} \tag{A6} \label{eq:A6}.
\end{equation}
\\
I get, for example,
\begin{align}
    [J_{a+, b+}, \ \Psi_\alpha] &= \Gamma_{a+} \{ \Gamma_{b+}, \ \Psi_\alpha \} - \{ \Gamma_{a+}, \ \Psi_\alpha \} \Gamma_{b+}, \notag \\
    [J_{a+, b-}, \ \Psi_\alpha] &= \Gamma_{a+} \{ \Gamma_{b-}, \ \Psi_\alpha \} - \{ \Gamma_{a+}, \ \Psi_\alpha \} \Gamma_{b-} \tag{A7} \label{eq:A7}.
\end{align}
Here, $\{ \Gamma_{a \pm}, \Psi_\alpha \}$ are given by
\begin{align}
    \{ \Gamma_{a \pm}, \ \Psi_\alpha \} &= \{ \Gamma_{a \pm}, (\Gamma_{0+})^{\alpha_{0} + 1/2} (\Gamma_{1+})^{\alpha_{1} + 1/2} \dots (\Gamma_{4+})^{\alpha_{4} + 1/2} \} \notag \\
    &= \{ \Gamma_{a \pm}, (\Gamma_{0+})^{\alpha_{0} + 1/2} \}  (\Gamma_{1+})^{\alpha_{1} + 1/2} \dots (\Gamma_{4+})^{\alpha_{4} + 1/2} \notag \\
    & \ \ \ \ - (\Gamma_{0 \pm})^{\alpha_0 + 1/2} [\Gamma_{a \pm}, (\Gamma_{1+})^{\alpha_{1} + 1/2} \dots (\Gamma_{4+})^{\alpha_{4} + 1/2}] \tag{A8} \label{eq:A8}.
\end{align}
The right hand side contains
\begin{align}
    [\Gamma_{a \pm}, (\Gamma_{1+})^{\alpha_{1} + 1/2} & \dots (\Gamma_{4+})^{\alpha_{4} + 1/2}] \notag \\
    &= \{ \Gamma_{a \pm}, (\Gamma_{1+})^{\alpha_{1} + 1/2} \} (\Gamma_{2+})^{\alpha_{2} + 1/2} \dots (\Gamma_{4+})^{\alpha_{4} + 1/2} \notag \\
    & \ \ \ \ - (\Gamma_{1+})^{\alpha_{1} + 1/2} \{ \Gamma_{a \pm}, (\Gamma_{2+})^{\alpha_{2} + 1/2} \dots (\Gamma_{4+})^{\alpha_{4} + 1/2} \} \notag.
\end{align}
The right hand side of this equation contains
\begin{align}
    \{ \Gamma_{a \pm}, (\Gamma_{2+})^{\alpha_{2} + 1/2} &\dots (\Gamma_{4+})^{\alpha_{4} + 1/2} \} \notag \\
    &= \{ \Gamma_{a \pm}, (\Gamma_{2+})^{\alpha_{2} + 1/2} \}  (\Gamma_{3+})^{\alpha_{3} + 1/2} (\Gamma_{4+})^{\alpha_{4} + 1/2}  \notag \\
    & \ \ \ \ - (\Gamma_{2+})^{\alpha_2 + 1/2} [\Gamma_{a \pm}, (\Gamma_{3+})^{\alpha_{3} + 1/2} (\Gamma_{4+})^{\alpha_{4} + 1/2}] \tag{A10} \label{eq:A10}.
\end{align}
And finally,
\begin{align}
    [\Gamma_{a \pm}, &(\Gamma_{3+})^{\alpha_{3} + 1/2} (\Gamma_{4+})^{\alpha_{4} + 1/2}] \notag \\
    &= \{ \Gamma_{a \pm}, (\Gamma_{3+})^{\alpha_{3} + 1/2} \} (\Gamma_{4+})^{\alpha_{4} + 1/2} - (\Gamma_{3+})^{\alpha_{3} + 1/2} \{ \Gamma_{a \pm}, (\Gamma_{4+})^{\alpha_{4} + 1/2} \} \notag.
\end{align}
Generally, I have
\begin{gather}
    \{ \Gamma_{a +}, (\Gamma_{b+})^{\alpha_{a} + 1/2} \} = 2 \delta_{\alpha, -1/2} \Gamma_{\alpha+}, \notag \\
    \{ \Gamma_{a-}, (\Gamma_{b+})^{\alpha + 1/2} \} = \delta_{\alpha, 1/2} \delta_{a, b} + 2 \delta_{\alpha, -1/2} \Gamma_{a-} \tag{A11} \label{eq:A11}.
\end{gather}
I start with 
\begin{equation}
    \alpha = \{ \alpha_{0}, \alpha_{1}, \alpha_{2}, \alpha_{3}, \alpha_{4} \} = \{ 1/2,1/2,1/2,1/2,1/2 \} \mbox{ that is } \Psi_{\alpha} = (\Gamma_{0+}) (\Gamma_{1+}) \dots (\Gamma_{4+}) \tag{A12} \label{eq:A12}.
\end{equation}
Obviously,
\begin{equation*}
    [J_{a+, b+}, \ \Psi_{\alpha}] = 0,
\end{equation*}
and 
\begin{align}
    [J_{a+, b-}, \ \Psi_{\alpha}] &= \Gamma_{a+} \{ \Gamma_{b-}, \Psi_\alpha \} = \delta_{ab} \ \Psi_\alpha \notag \\
    [J_{a-, b+}, \ \Psi_{\alpha}] &= \Gamma_{a-} \{ \Gamma_{b+}, \Psi_\alpha \} - \{ \Gamma_{a-}, \Psi_\alpha \} \Gamma_{b+} = - \{ \Gamma_{a-}, \Psi_\alpha \} \Gamma_{b+} \notag \\
    &= - \delta_{ab} \ \Psi_\alpha \notag \\
    [J_{a-, b-}, \ \Psi_{\alpha}] &= \Gamma_{a-} \{ \Gamma_{b-}, \Psi_\alpha \} - \{ \Gamma_{a-}, \Psi_\alpha \} \Gamma_{b-} = \Gamma_{a-} \Gamma_{b-} \Psi_\alpha \notag \\
    & = (-1)^b (-1)^{a+\theta (a-b)} \ \Psi_{\alpha \{ a-, \ b- \}} \tag{A13} \label{eq:A13}.
\end{align}
Here, the notation $\alpha \{a-, b-\}$ means $\alpha = \{\alpha_{0}, \alpha_{1}, \alpha_{2}, \alpha_{3}, \alpha_{4}\} $ that with $\alpha_{a} \longrightarrow \alpha_{a} - 1/2$, $\alpha_{b} - 1/2$ $\alpha_{b} \longrightarrow \alpha_{b} - 1/2$, and other $\alpha$'s unchanged.  
Therefore $\Psi$ vanishes unless both $\alpha_{a} = 1/2$ and $\alpha_{b} = 1/2$.  
$\alpha \{a+, b+\} $ can be defined similarly.  
$\Psi$ also vanishes if $a=b$.  
\\
Generally,
\begin{align}
    &[J_{a+, b-}, \ \Psi_{\alpha}] = \Gamma_{a+} \Gamma_{b+} \Psi_\alpha - \Psi_\alpha \Gamma_{a+} \Gamma_{b+} \notag \\
    & \ \ = \{ (-1)^{N_\alpha (b) + N_{\alpha (b+)} (a)} - (-1)^{M_a (a) + M_{\alpha (a+)} (b)} \} \ \delta_{\alpha_a + 1/2, 0} \ \delta_{\alpha_b + 1/2, 0} \ \Psi_{\alpha (a+, b+)} \tag{A14} \label{eq:A14}.
\end{align}
Here, $N_{\alpha} (a)$ indicates the number of creation operators in the state $\alpha$ sitting on the left of the creation operator $a+$, and $M_{\alpha} (a)$ means the number of creation operators in the state $\alpha$ sitting on the right of the creation operator $a+$.  
$\alpha (a+)$ is the state with $\alpha = \{ \alpha_{0}, \alpha_{1}, \alpha_{3}, \alpha_{4}\} $, $\alpha_{a} \longrightarrow \alpha_{a} + 1/2$ and other $\alpha$'s are unchanged.  
\begin{align}
    [J_{a-, b+}, \Psi_{\alpha}] &= \Gamma_{a-} \Gamma_{b+} \Psi_\alpha -  \Psi_\alpha \Gamma_{a-} \Gamma_{b+} \notag \\
    &= (-1)^{N_\alpha (b) + N_{\alpha (b+)} (a)} \delta_{\alpha_a - 1/2, \ 0} \delta_{\alpha_b + 1/2, \ 0} \Psi_{\alpha (a-, \ b+)} - \delta_{a,b} \Psi_\alpha \notag \\
    [J_{a+, b-}, \Psi_{\alpha}] &= \Gamma_{a+} \Gamma_{b-} \Psi_\alpha -  \Psi_\alpha \Gamma_{a+} \Gamma_{b-} \notag \\
    &= (-1)^{N_\alpha (b) + N_{\alpha (b+)} (a)} \delta_{\alpha_a + 1/2, \ 0} \delta_{\alpha_b - 1/2, \ 0} \Psi_{\alpha (a+ \ b-)} + \delta_{a,b} \Psi_\alpha \notag \\
    [J_{a-, b-}, \Psi_{\alpha}] &= \Gamma_{a-} \Gamma_{b-} \Psi_\alpha = (-1)^{N_\alpha (b) + N_{\alpha (b-)} (a)} \delta_{\alpha_a - 1/2, \ 0} \delta_{\alpha_b - 1/2, \ 0} \Psi_{\alpha (a-, \ b-)} \tag{A15} \label{eq:A15}.
\end{align}
So far I have been discussing $SO(9,1)$ generators.  
To go to $SO(10,1)$ I must add, $J_{10, a\pm}$ and $J_{a\pm, 10}$ which are given by $\Gamma_{10} \Gamma_{a+}$ and $\Gamma_{a+} \Gamma_{10}$ respectively, with $\Gamma_{10} = \Gamma_{0} \Gamma_{1} \dots \Gamma_{9} = 32 \ S_{0} S_{1} S_{2} S_{3} S_{4}$.  
Here, $S_{a} = \Gamma_{a+} \Gamma_{a-} - 1/2$.  
\begin{align}
    [J_{10, a+}, \ \Psi_\alpha] &= \Gamma_{10} \Gamma_{a+} \Psi_\alpha - \Psi_\alpha \Gamma_{10} \Gamma_{a+} = \Gamma_{10} \Gamma_{a+} \Psi_\alpha \notag \\
    &= \{ (-1)^{N_c (a)+1} (-1)^{N_a (a)} + (-1)^{M_\alpha (a)} \} \delta_{\alpha_a + 1/2, \ 0} \Psi_{\alpha (a+)} \notag \\
    [J_{10, a-}, \Psi_\alpha] &= \Gamma_{10} \Gamma_{a-} \Psi_\alpha - \Psi_\alpha \Gamma_{10} \Gamma_{a-} = \Gamma_{10} \Gamma_{a-} \Psi_\alpha \notag \\
    &= (-1)^{N_c (a)+1} (-1)^{N_a (a)} \delta_{\alpha_a - 1/2, \ 0} \Psi_{\alpha (a-)} \tag{A16} \label{eq:A16}.
\end{align} 
In summary,
\begin{align}
    \chi_\alpha^{\alpha (a+, b+)}; &\ a+, \ b+ \notag \\
   & =  \{ (-1)^{N_\alpha (b) + N_{\alpha (b+)} (a)} - (-1)^{M_\alpha (a) + M_{\alpha (a+)} (b)} \} \ \delta_{\alpha_a + 1/2, \ 0} \ \delta_{\alpha_b + 1/2, \ 0} \notag \\
    \chi_\alpha^{\alpha (a-, b+)}; &\ a-, \ b+ \notag \\
    &=  (-1)^{N_\alpha (b) + N_{\alpha (b+)} (a)} \ \delta_{\alpha_a - 1/2, \ 0} \ \delta_{\alpha_b + 1/2, \ 0}, \ (a \ne b), \ \chi_\alpha^\alpha; a-, a+ = -1 \notag \\
    \chi_\alpha^{\alpha (a+, b-)}; &\ a+, \ b- \notag \\
    &=  (-1)^{N_\alpha (b) + N_{\alpha (b+)} (a)} \ \delta_{\alpha_a + 1/2, \ 0} \ \delta_{\alpha_b - 1/2, \ 0} \ \Psi_{\alpha (a+, b-)}, \ (a \ne b),  \ \chi_\alpha^\alpha; a+, a- = -1 \notag \\
    \chi_\alpha^{\alpha (a-, b-)}; &\ a-, \ b- =  (-1)^{N_\alpha (b) + N_{\alpha (b-)} (a)} \ \delta_{\alpha_a - 1/2, \ 0} \ \delta_{\alpha_b - 1/2, \ 0} \notag \\
    \chi_\alpha^{\alpha (a+)}; \ 1&0, \ a+ =  \{ (-1)^{N_c (a)+1} (-1)^{N_\alpha (a)} + (-1)^{M_\alpha (a)} \delta_{\alpha_a + 1/2, \ 0} \notag \\
    \chi_\alpha^{\alpha (a-)}; \ 1&0, \ a- =  (-1)^{N_\alpha (b)+N_{\alpha (b-)} (a)} \delta_{\alpha_a - 1/2, \ 0} \delta_{\alpha_b - 1/2, \ 0} \tag{A17} \label{eq:A17}.
\end{align}



\begin{thebibliography}{999}
\bibitem{West}
  P. West, Class. Quant. Grav. 18 (2001) 4443, hep-th/0104081. \\
  P. West, Phys. Lett. B 575 (2003) 333-342, hep-th/0307098. \\
  A. Tumanov and P. West, Phys. Lett. B759 (2016) 663, arXiv: 1512.01644. \\
  A. Tumanov and P. West, Phys. Lett. B758 (2016) 278, arXiv: 1601.03974. \\
  P. West, A brief review of E theory, Proceedings of Abdus Salam's 90th Birthday meeting, 
  25-28 January 2016, NTU, Singapore, Editors L. Brink, M. Duff and K. Phua, 
  World Scientific Publishing and IJMPA, Vol 31, No 26 (2016) 1630043. 
\bibitem{Cremmer}
  E. Cremmer, B. Julia and J. Scherk, Phys. Lett. 76B (1978) 409.
\bibitem{HS}
  H. Sugawara, Phys. Rev. 170 (1968) 1659.
\bibitem{Kroll}
  N.M. Kroll, T.D. Lee and B. Zumino, Phys. Rev. 157 (1967) 1376.
\bibitem{Schwinger}
  J. Schwinger, Phys. Rev. 130 (1963) 406; ibid. 800.
\bibitem{Kac}
  V.G. Kac, 
  \textit{``Infinite Dimensional Lie Algebra-An Introduction,''} 
  Progress in Mathematics, Edited by J. Coates and S. Helgason, Vol. 44 (1983).
\bibitem{Serre}
  J.P. Serre, 
  \textit{``Complex Simple Lie Algebra,''} 
  Springer (2001).
\bibitem{West2}
  P. West, 
  \textit{``Introduction to Strings and Branes,''} 
  Cambridge (2012).
\bibitem{HSandMY}
  H. Sugawara and M. Yoshimura, Phys. Rev. 173 (1968) 1419.
\end{thebibliography}
\end{document}